\documentclass[twocolumn,showpacs,preprintnumbers,amsmath,amssymb]{revtex4}
\begin{document}

\title{Superspace formulations of the (super)twistor string}

\author{
Igor A. Bandos$^{\dagger}$$^{\ast}$, Jos\'e A. de
Azc\'arraga$^{\dagger}$ and C\`{e}sar Miquel-Espanya$^{\dagger}$}

\bigskip

\address{$^{\dagger}$Departamento de
F\'{\i}sica Te\'orica, Univ.~de Valencia and IFIC (CSIC-UVEG), 46100-Burjassot
(Valencia), Spain
\\
$^{\ast}$Institute for Theoretical Physics, NSC ``Kharkov Institute of
Physics  and Technology'',  UA61108, Kharkov, Ukraine}

\def\theequation{\arabic{section}.\arabic{equation}}

\begin{abstract}
The superspace formulation of the worldvolume action of twistor
string models is considered. It is shown that for the
Berkovits-Siegel closed twistor string such a formulation is
provided by a N=4 twistor-like action of the tensionless
superstring.  A similar inverse twistor transform of the open
twistor string model (Berkovits model) results in a dynamical system
containing two copies of the $D=4$, $N=4$ superspace coordinate
functions, one left-moving and one right-moving, that are glued by
the boundary conditions.

We also discuss possible candidates for a tensionful superstring
action leading to the twistor string in the tensionless limit as
well as multidimensional counterparts of twistor strings in the
framework of both `standard' superspace and superspace enlarged by
tensorial coordinates (tensorial superspaces), which constitute a
natural framework for massless higher spin theories.

\end{abstract}

\pacs{11.30.Pb, 11.25.-w, 11.10.Kk, 12.60Jv; \\
  FTUV-06-0405 , IFIC/06-06 ; $\;$ April 5, 2006}

\maketitle

\section{Introduction}

The connection between Yang-Mills and string theories was
reconsidered in \cite{Witten03} using the twistor approach
\cite{Pen} (see \cite{Nair88} for an earlier study). It was
originally noticed that a class of perturbative tree amplitudes
for the gauge fields of $N=4$ supersymmetric Yang-Mills theories
were reproduced from a string moving in the projective
$\mathbb{CP}^{(3|4)}$ superspace. $\mathbb{CP}^{(3|4)}$ is a
Calabi-Yau supermanifold, the bosonic body of which,
$\mathbb{CP}^3$, is Penrose twistor space \cite{Pen}.

The class of Yang-Mills amplitudes that may be described by the
twistor string model \cite{Witten03} was then extended in
\cite{RSV04,W+04}. In particular, one loop amplitudes are also
amenable to a twistor presentation using a technique
\cite{W+04,W+CS04,Bena04} suggested by the twistor string
approach. This further supports the original idea of the existence
of a deep connection between the supertwistor string and $N=4$
super-Yang-Mills gauge theories in the usual $D=4$
(super)spacetime. The restrictions on a possible full
identification of these models come from the impossibility of
isolating the closed string sector of the twistor string from the
open one, and from the observation that the closed string sector,
in view of its conformal invariance, should lead to conformal
supergravity \cite{B+W-2004}, which has itself problems for its
physical treatment. Recently \cite{Mason05}, a twistor string-like
generating functional for super-Yang-Mills amplitudes was derived
from a twistor reformulation of the super-Yang-Mills action which,
in turn, followed from its `asymmetric' formulation
\cite{Chalmers+Siegel96}. An analogous formulation for gravity
that might be related with the closed twistor string was
considered in \cite{Hull+2005}.

At the same time, alternative twistor string models were proposed
in \cite{NB04,Vafa04} and \cite{Siegel04}. It was argued in
\cite{Vafa04} that the twistor string might be related with the
${\cal N}=2$ spinning string, which has ${\cal N}=2$ extended
worldvolume supersymmetry and is defined in a $D=4$ spacetime with
two time-like directions (see \cite{LePo04} for further
discussion). The problem of relating the twistor string with
two-time physics was considered in \cite{Bars04} where possible
higher dimensional generalizations were also discussed.

To look for a (super)spacetime formulation of the twistor string
that originally had been given only in terms of supertwistor
variables but not in the usual spacetime or superspace
coordinates, Siegel proposed a new twistor string action
\cite{Siegel04} in terms of the Atiyah-Drinfel'd-Hitchin-Manin
(ADHM) \cite{AHDM:78} (super)twistors (see \cite{Siegel94}).

One of the messages of this paper is that the standard (not ADHM)
twistor superstring action, at least the closed string chiral
version of Siegel \cite{Siegel04}, can be rewritten in terms of
superspace coordinates in a different manner using a method
similar to the change of variables that relates the different
forms of the Ferber-Shirafuji superparticle action \cite{FS78-83}
(see also \cite{BBCLGS}). The spacetime/superspace action that is
classically equivalent to the twistor superstring \cite{Siegel04}
has 8 $\kappa$-symmetries and turns out to be a straightforward
$N=4$ generalization of the $D=4$ tensionless superstring action
in \cite{BZnull,BZnullr}.

A similar twistor transform of the Berkovits model for the open
twistor string results in an action formulated in terms of two
copies of the coordinate functions of $D=4$, $N=4$ superspace. We
notice in passing that such a set of variables, albeit for $N=1$,
was used in  \cite{Ivanov+Isaev88} to write an equivalent form of
the $N=2$ Green-Schwarz superstring action.

The way of quantizing the twistor string discussed in
\cite{Witten03,NB04,Siegel04,B+W-2004} makes it clear that this
string should be understood as the tensionless limit of a
tensionful string model. The reason is that in the case of the
intrinsically tensionless superstring (`null-superstring') its
conformal invariance is maintained by a continuous mass spectrum
\cite{BZnullb,BZnull}, while the tensionless limit of a tensionful
string rather contains a set of massless fields
\cite{Gross,Lindstrom03,Bonelli}. The relation of such a
tensionful model with the so-called QCD string \cite{Siegel-QCD}
was discussed in \cite{Siegel04}, but in a purely bosonic context.
The correspondence with the null-superstring of
\cite{BZnull,BZnullr} makes transparent that the $N$=1 ($N$=2)
counterparts of the twistor string action appear in the
tensionless limit of the standard $N=1$ ($N=2$) Green-Schwarz
superstring models in $D$=4. The search for a tensionful parent
action for the standard (super-Yang-Mills related) $N$=4 twistor
string will lead us naturally to enlarge the $D=4$, $N=4$
$\Sigma^{(4|4N)}$ superspace to the $D=10$, $N=1$ superspace
$\Sigma^{(10|16)}$ or to a tensorial superspace
$\widetilde{\Sigma}^{(4+6|16)}$ with additional antisymmetric
tensor coordinates.

\section{Supertwistor string models}

\subsection{Siegel's closed string action}\label{subsect.Siegel's Action}

To our knowledge, there are at present four versions of the
supertwistor string action, that of Witten \cite{Witten03}, a
constrained sigma model the tangent superspace of which is
$\mathbb{CP}^{(3|4)}$, the one put forward by Berkovits
\cite{NB04} involving {\it two} supertwistors, and the two
proposed by Siegel in \cite{Siegel04}.

The simplest action is that of the closed twistor string model of
the first part of ref. \cite{Siegel04} (we do not discuss here the
second, ADHM twistor action given in \cite{Siegel04}, which
includes explicitly the spacetime coordinates and that in this
sense `untwists'  the twistor superstring). It is given by
\begin{eqnarray}\label{TWS-S}
S &=& \int_{W^2} e^{++} \wedge \Bar{\Upsilon}_\Sigma\;
\nabla\Upsilon^\Sigma \; + d^2\xi L_{G} = \qquad \nonumber
\\ &=& \int d^2 \xi \, [ \sqrt{|\gamma(\xi) |} \,\; \Bar{\Upsilon}_\Sigma(\xi)\,
\nabla_{\!_{--}} \Upsilon^\Sigma(\xi) \;  + L_{G} ] \; , \qquad
\end{eqnarray}
where $e^{\pm\pm}= d\xi^m e_m^{\pm\pm}(\xi)$ are the worldsheet
zweibein one-forms and $e^{++}\wedge e^{--} = d^2 \xi \,
\sqrt{|\gamma |} $ is the invariant surface element of the
worldsheet $W^2$. The basic worldsheet fields
\begin{eqnarray}\label{YTW}
 && \Upsilon^\Sigma =  ( \mu^{\dot{\alpha}} \, , \, \lambda_\alpha \, ;
\, \eta_i ) =  \Upsilon^\Sigma (\xi) \; , \qquad
\\ \nonumber \alpha =1, 2 \; & , & \qquad \dot{\alpha}=1,2\; , \qquad
i=1,2,3,4\; ,
\end{eqnarray}
determine the $(N=4)$ Ferber supertwistor \cite{FS78-83}
$\Upsilon^\Sigma$ and
\begin{eqnarray}\label{bYTW} \Bar{\Upsilon}_\Sigma :=
 \left(\Upsilon^\Pi\right)^\ast \Omega_{\Pi\Sigma} = (
\bar{\lambda}_{\dot{\alpha}} \,  , \, - \bar{\mu}^\alpha \,  ; \,
2i \bar{\eta}^i ) \;
, \qquad
\\ \nonumber \alpha =1, 2 \; , \qquad \dot{\alpha}=1,2\; , \qquad
i=1,2,3,4
\end{eqnarray}
\footnote{Following \cite{Witten03} we use the convention of
adding a bar to all the spinors in $\Bar{\Upsilon}_\Sigma$ rather
than associating the bar to the dotted ones ({\it e.g.}, by
denoting ($\lambda_\alpha\, , \, \bar{\mu}^{\dot{\alpha}}$) the
two Weyl spinors in $\Upsilon^\Sigma$ that make up a Dirac
spinor).} is defined through the $SU(2,2|4)$-invariant tensor
\begin{eqnarray}\label{OmSU224}
\Omega_{\Sigma\Pi}  = \left( \begin{matrix}
 0 & -\delta_{\dot{\alpha}}{}^{\dot{\beta}} & 0 \cr
\delta^{\alpha}{}_{\beta} & 0 & 0 \cr 0 & 0 &  2i \cr
\end{matrix}\right)\;\; . \quad
\end{eqnarray}
Finally,
\begin{eqnarray} \label{nabla=} \nabla = e^{++}\nabla_{++} +
e^{--}\nabla_{--}=d- iB\;
\end{eqnarray}
is the worldsheet covariant derivative with the $U(1)$-connection
$B$. In (\ref{TWS-S}), $L_{G}$ is the action for the worldsheet
fields that are used to construct the Yang-Mills symmetry current
(one can use {\it e.g.}, the worldsheet fermionic degrees of
freedom, as discussed below).

The target supermanifold $\mathbb{C}\mathbb{P}^{(3|4)}$ of the
$N=4$ supertwistors (\ref{YTW}), which generalize the Penrose
twistors \cite{Pen}, defines a fundamental representation of the
$SU(2,2|4)$ superconformal group ($SU(2,2) \sim SO(2,4)$). Thus,
the action (\ref{S-0S}) is superconformally ($SU(2,2|4)$-)
invariant by construction. Such an action can be also written for
$N\not=4$ supertwistors; in this case it possesses $SU(2,2|N)$
superconformal symmetry, but $\mathbb{CP}^{(d|N)}$ is a Calabi-Yau
manifold only if $d+1=N$. The set of special properties of the
$N=4$ case includes the existence of the holomorphic integral
measure on $\mathbb{CP}^{(3|4)}$,
\begin{eqnarray} \label{Om3-4}
& \Omega_{(3|4)} =  \;  \Omega_{(3|0)} \; \epsilon_{ijkl} {\partial\over
\partial \eta_i}  \; {\partial\over
\partial \eta_j}  \; {\partial\over \partial \eta_k} \; {\partial\over \partial
\eta_l}\; , \qquad \nonumber \\
& \Omega_{(3|0)} =  \epsilon_{\alpha'\beta'\gamma'\delta'}
\Upsilon^{\alpha'} d\Upsilon^{\beta'} \wedge d\Upsilon^{\gamma'}
\wedge d\Upsilon^{\delta'} \; ,
 \qquad
\end{eqnarray}
where $\alpha'=1,\dots,4\ (=1,2,\dot 1,\dot 2)$ . The
$\Omega_{(3|4)}$ integral form is invariant under the $U(1)$-phase
transformations of the twistors,
\begin{eqnarray} \label{U(1)Y}
\Upsilon^\Sigma \mapsto e^{i\beta} \Upsilon^\Sigma  \; ,
\qquad \bar{\Upsilon}_\Sigma
\mapsto e^{-i\beta} \bar{\Upsilon}_\Sigma , \; \qquad
\end{eqnarray}
and also under the scaling
\begin{equation}\label{scaling}
\Upsilon^\Sigma \mapsto e^{\beta^\prime}\Upsilon^\Sigma\quad .
\end{equation}
These two transformations are also symmetries of the action
(\ref{TWS-S}) provided the scaling of the twistors, Eq.
(\ref{scaling}), is supplemented by the scaling of the vielbein form
\begin{equation}
\label{Weyl}
 e^{++}\mapsto e^{-2\beta'}e^{++}\quad.
\end{equation}

Because of the covariant derivative (\ref{nabla=}), the $U(1)$ gauge
transformations (\ref{U(1)Y}) (now with local parameter $\beta(\xi)$
under which $B \mapsto B+ d\beta$) are a gauge symmetry of the
action (\ref{TWS-S}). On the other hand, the r\^ole of this $U(1)$
connection $B$ is analogous (as noted in \cite{Siegel04}) to that of
the auxiliary worldsheet metric in the standard superstring model.
Namely, its equations of motion impose on the supertwistor
$\Upsilon$ the constraint
\begin{eqnarray}\label{YY=0}
\Bar{\Upsilon}_\Sigma\, {\Upsilon}^\Sigma = \bar{\lambda}_{\dot{\alpha}} \,
{\mu}^{\dot{\alpha}} - \bar{\mu}^{{\alpha}} {\lambda}_{{\alpha}} +2i \bar{\eta}^i
\eta_i =0 \;
\end{eqnarray}
which, in the Hamiltonian framework, is the generator of the $U(1)$
symmetry (\ref{U(1)Y}).

As the constraint (\ref{YY=0}) appears as a non-dynamical equation
of motion for the auxiliary field $B$, one can consider the action
\begin{eqnarray}\label{TWS-S1}
S&=& \int e^{++} \wedge \Bar{\Upsilon}_\Sigma\; d\Upsilon^\Sigma   + d^2\xi L_{G} =
\qquad \\  \nonumber  &=& \int e^{++} \wedge (\bar{\lambda}_{\dot{\alpha}} \,
d{\mu}^{\dot{\alpha}} - \bar{\mu}^{{\alpha}} d{\lambda}_{{\alpha}} +2i \bar{\eta}^i
d\eta_i) + d^2\xi L_{G} \; ,
\end{eqnarray}
where the supertwistor variables are constrained by (\ref{YY=0}),
as an alternative to (\ref{TWS-S}). In this form the action does
not contain the connection $B$, but the $U(1)$ gauge symmetry
still holds due to the constraint (\ref{YY=0}). This constraint
also makes the action (\ref{TWS-S1}) invariant under the {\it
local} worldsheet Lorentz $SO(1,1)$, which here is equivalent to
the scaling or local $GL(1,\mathbb{R})$) symmetry (\ref{scaling})
\footnote{The worldsheet local Lorentz $SO(1,1)$ invariance of
(\ref{TWS-S}) for unconstrained supertwistors holds if $B$ has
nontrivial $SO(1,1)$ transformation properties $B\mapsto B +
id\alpha$, so that $B$ is no longer a real connection ({\it cf.}
\cite{NB04,Siegel04}). Both local symmetries can be maintained if
$B$ is taken to be a complex $GL(1,\mathbb{C})$ connection. The
reality of the action with complex $B$ would hold if (\ref{TWS-S})
were written in the `symmetric' form: $\Bar{\Upsilon}_\Sigma\;
\nabla\Upsilon^\Sigma \; \mapsto \; {1\over
2}(\Bar{\Upsilon}_\Sigma\; \nabla\Upsilon^\Sigma -
\nabla\Bar{\Upsilon}_\Sigma\; \Upsilon^\Sigma )$. For the action
(\ref{TWS-S}) with a $U(1)$ connection $B$ this holds, up to
boundary contributions, provided that
$B_{--}^\ast = B_{--} + {i\over 2} e_{--}{}^m \nabla_{++}e_m^{++}
- {i\over 2} e_{++}{}^m \nabla_{--}e_m^{++}$ ($B_{++}$ does not
enter the action (\ref{TWS-S}) ). }.

\subsection{On the Yang-Mills current part of the action}
\label{subsect.YM current part}

The simplest choice for $L_{G}$ in (\ref{TWS-S}) is the free
fermion action which allows one to construct the current ${\cal
J}^r= \bar{\psi}_J T^{rJ}_I\psi^I$ for the Yang-Mills gauge group
$G$ (where $T^{r}{}^{\;J}_I$ is the matrix representation for its
infinitesimal generators) and, hence, to describe the coupling of
the string to the Yang-Mills gauge field, $\int tr{\cal J}{\cal
A}\propto \int {\cal J}^r{\cal A}^r$, according to
\cite{Halpern,Windey,Sagnotti87}. In the action (\ref{TWS-S}) for
the closed string \cite{Siegel04}, these free fermions should have
the same two-dimensional (worldsheet) `chirality' as the
supertwistor variables \cite{WS05}. Thus the Lagrangian including
the vertex operator reads
\begin{eqnarray}\label{LYM-ff+}
d^2\xi \left[L_{G} + \hbox{\rm tr}( {\cal J}{\cal A})\right]=
{1\over 2} e^{++} \wedge (\bar{\psi}_{I} D\psi^I - D\bar{\psi}_{I}
\, \psi^I) \;
,\nonumber\\
\end{eqnarray}
where $D$ is the Yang-Mills covariant derivative $D\psi^I=d\psi^I +
{\cal A}^I{}_J\psi^J$ and $\bar{\psi}_I= (\psi^J)^*{\cal C}_{J^*I}$
with ${\cal C}_{J^*I}$ being invariant under the gauge group $G$;
for instance ${\cal C}_{J^*I}= {\delta}_{J^*I}$ for $G=U(m)$.

\subsection{Berkovits's open string action}

Berkovits's  open string version of the twistor string \cite{NB04}
contains two supertwistor fields, a left moving $\Upsilon^{-\Sigma
}$ and a right moving $\Upsilon^{+\Sigma }$,
\begin{eqnarray}\label{YTWL-R}
\Upsilon^{-\Sigma } = (\mu^{-\dot{\alpha}} \, , \,
\lambda^-_{\alpha} \,
 ; \, \eta^-_{i} ) \; , \qquad \nonumber
\\ \Upsilon^{+\Sigma } = (\mu^{+\dot{\alpha}} \, , \, \lambda^+_{\alpha} \,
 ; \, \eta^+_{i} ) \; . \qquad
\end{eqnarray}
The worldsheet action reads
\begin{eqnarray}\label{TWS-B}
S &=& \int_{W^2} e^{++} \wedge \Bar{\Upsilon}^-_{\Sigma } \nabla (\Upsilon^{-\Sigma} )
- e^{--} \wedge \Bar{\Upsilon}^+_{\Sigma} \nabla (\Upsilon^{+
\Sigma} )   + \quad \nonumber \\
&& + \int_{W^2} d^2\xi L_{G} \; ,
\end{eqnarray}
where $L_{G}$ is the Lagrangian for the YM current ({\it e.g.}
${1\over 2} e^{++} \wedge (\bar{\psi}_{+I} d\psi_+^I -
d\bar{\psi}_{+I} \, \psi_+^I) + {1\over 2} e^{--} \wedge
(\bar{\psi}_{-I} d\psi_-^I - d\bar{\psi}_{-I} \, \psi_-^I)$ for
the left and right fermionic fields $\psi_-,\ \psi_+$). The above
notation explains why the worldsheet supervielbein forms were
denoted by $e^{++}$ and $e^{--}$: the double sign superscript
indicates $SO(1,1)$-vector transformation properties
$(e^{\pm\pm}\rightarrow e^{\mp2\beta'}e^{\pm\pm})$, while the
single $\pm$ superscripts were reserved for supertwistors
$\Upsilon^{\pm\Sigma}$ to indicate their spinor
$(\Upsilon^{\pm}\rightarrow e^{\mp\beta'}\Upsilon^\pm)$
transformation properties under the worldsheet Lorentz group.

The action (\ref{TWS-B}) assumes boundary conditions that
identify, in particular, the left and the right supertwistors
\footnote{In the original paper \cite{NB04}
$\Bar{\Upsilon}^-_{\Sigma }$, like in (\ref{TWS-B}), is the
canonically conjugate momentum of $\Upsilon^{-\Sigma}$ but, unlike
in (\ref{TWS-B}), it is not its complex conjugate; the complex
conjugate of $\Upsilon^{-\Sigma}$ is there identified with
$\Upsilon^{+ \Sigma}$ ($\overline{\Upsilon^{-\Sigma}}=
\Upsilon^{+\Sigma}$ in \cite{NB04}). We, however, take the point
of view of \cite{Siegel04,Bars04}, in which, like in the standard
quantization of supertwistors \cite{FS78-83}, the canonically
conjugate supertwistors basically coincide with their complex
conjugates.} on the worldsurface boundary $\partial W^2$:
\begin{eqnarray}\label{TWS-Bbc}
\Upsilon^{-\Sigma} \vert_{\partial W^2}=
\Upsilon^{+\Sigma}\vert_{\partial W^2}\; ,  \qquad
\bar{\Upsilon}^-_{ \Sigma} \vert_{\partial W^2}=
\bar{\Upsilon}^+_{\Sigma}\vert_{\partial W^2}\; , \qquad
\end{eqnarray}
as well as the left and right currents. Siegel \cite{Siegel04}
motivated his modification (\ref{TWS-S}) of the Berkovits twistor
string (\ref{TWS-B}) by stating that the boundary conditions do
not play any r\^ole except that of halving the number of the
twistor degrees of freedom. Specifically, the identification
(\ref{TWS-Bbc}) of the two supertwistors on the boundary $\partial
W^2$ of $W^2$ allows one to construct all the open string vertex
operators using only one set of twistor variables, either
$(\Upsilon^{-\Sigma}\, , \, \bar{\Upsilon}^-_{\Sigma})$ or
$(\Upsilon^{+\Sigma}\, ,  \, \bar{\Upsilon}^+_{\Sigma})$. It was
also noticed that the closed string version is more natural for a
spacetime interpretation, which was constructed by moving from the
Penrose-twistor string to an alternative, `six dimensional'
ADHM-twistor string \cite{Siegel04}. We will present here another,
more straightforward way to arrive at the spacetime or, more
precisely, at the standard superspace presentation.

\subsection{Witten's action}

For completeness we describe here the original Witten's proposal
\cite{Witten03} for a $\mathbb{CP}^{(3|4)}$ twistor string. It uses
only one supertwistor and it is based on the following action
\begin{eqnarray}\label{TWS-W}
& S_W & = \int_{W^2} \left[  \nabla \Bar{\Upsilon}_\Sigma \wedge
\ast \nabla \Upsilon^\Sigma \,
 + d^2\xi \; \Xi(\xi) \left( \Bar{\Upsilon}_\Sigma\Upsilon^\Sigma
 - r \right) \right]  = \nonumber \\
&=& \int  e^{++} \wedge e^{--} \; \left[ \nabla_{\!\!_{++}} \Bar{\Upsilon}_\Sigma \;
\nabla_{\!\!_{--}} {\Upsilon}^\Sigma + \nabla_{\!\!_{--}} \Bar{\Upsilon}_\Sigma \;
\nabla_{\!\!_{++}} {\Upsilon}^\Sigma \right]\nonumber \\
  && + \int d^2\xi \Xi(\xi) \left( \Bar{\Upsilon}_\Sigma  \Upsilon^\Sigma
 - r \right)
 \;
\end{eqnarray}
describing a $\mathbb{CP}^{(3|4)}$ sigma-model subject to the
additional constraint
\begin{eqnarray}\label{YY=r}
\Bar{\Upsilon}_\Sigma {\Upsilon}^\Sigma = r
\end{eqnarray}
for some constant $r$, introduced into the action through the
Lagrange multiplier $\Xi(\xi)$. In Eq. (\ref{TWS-W}), $\ast$ is
the Hodge operator for the auxiliary worldsheet metric,
\begin{eqnarray}\label{Hodge*}
\ast e^{--}= e^{--}\; , \quad \ast e^{++}=  - e^{++}\; .
\end{eqnarray}

In the case of particle mechanics \cite{FS78-83} the modification
of the twistor constraint (\ref{YY=0}) by a nonvanishing $r$ (Eq.
(\ref{YY=r})) is known to describe a massless particle with
helicity $s=r/2$. On the other hand, it is also known that, due to
the noncommutativity of ${\Upsilon}^\Sigma$ and
$\Bar{\Upsilon}_\Sigma$ in a quantum description, the classical
constraint (\ref{YY=0}) can also lead after quantization to
(\ref{YY=r}) with a nonvanishing $r$ \footnote{In the case of
particle mechanics, the quantization of $r=2s$ in units of $\hbar$
can easily be obtained as the requirement that the wave function
be well defined as a function of complex variable {\it i.e.}, that
under a $2\pi$ phase transformation of the bosonic spinor argument
the phase of the wave function is shifted by $kr\pi$, $k \in
\mathbb{Z}$.
See references  and discussion in \cite{B90}.\\
For a two-twistor description of massive particles see
\cite{BALM04} and references therein; a one-twistor description
has recently been developed in \cite{BaPi06}. }.

The covariant derivative $\nabla$ in (\ref{TWS-W}) contains the
U(1)-gauge field $B$:  $\; \nabla \Upsilon_{\Sigma}=
d\Upsilon_{\Sigma} - i B\, \Upsilon_{\Sigma}$, Eq. (\ref{nabla=}).
Due to the constraint (\ref{YY=r}), the equations of motion for
the gauge field $B$ can be written as
\begin{eqnarray}\label{YdY=}
0= \Bar{\Upsilon}_\Sigma \nabla {\Upsilon}^\Sigma  &=&
\Bar{\Upsilon}_\Sigma  d{\Upsilon}^\Sigma - i B \;
\Bar{\Upsilon}_\Sigma {\Upsilon}^\Sigma \,  = \nonumber
\\ &=& d{\Upsilon}^\Sigma \; \Bar{\Upsilon}_\Sigma - i B\; r
  \; . \qquad
\end{eqnarray}
Hence for a nonvanishing $r$ the U(1)-gauge field may be expressed
in terms of the supertwistor and its conjugate,
\begin{eqnarray}\label{B=}
B= - {i\over r}   \overline{{\Upsilon}}_\Sigma d{\Upsilon}^\Sigma \; .
\end{eqnarray}
In this case the Lagrange multiplier $\Xi(\xi)$ is also expressed in
terms of the twistor by the solution
\begin{eqnarray}\label{Xi=}
d^2\xi \, \Xi= + {1\over r} \nabla \overline{{\Upsilon}}_\Sigma \wedge * \nabla
{\Upsilon}^\Sigma \;
\end{eqnarray}
of the supertwistors equations of motion
\begin{eqnarray}\label{ddY=XiY}
&& \nabla * \nabla {\Upsilon}^\Sigma = d^2\xi \, \Xi {\Upsilon}^\Sigma \; , \nonumber \\
&& \nabla *  \nabla \overline{{\Upsilon}}_\Sigma = d^2\xi \, \Xi
\overline{{\Upsilon}}_\Sigma \; .
\end{eqnarray}
[To arrive at (\ref{Xi=}) one uses the constraint (\ref{YY=r}) and
equations (\ref{YdY=})]. Notice that inserting (\ref{Xi=}) back into
the equations of motion (\ref{ddY=XiY}) one finds
\begin{eqnarray}\label{DDY=YdYdY}
&& \nabla * \nabla {\Upsilon}^\Sigma = - {1\over r} {\Upsilon}^\Sigma \; \nabla
\overline{\Upsilon}_\Pi \wedge  * \nabla {\Upsilon}^\Pi
 \;
\end{eqnarray}
and its c.c. expression. The {\it r.h.s.} is proportional to the
trace of the energy-momentum tensor $T_{mn} \propto  \nabla_m
\overline{\Upsilon}_\Pi \;\nabla_n {\Upsilon}^\Pi$. This vanishes if
the equation of motion for the auxiliary worldvolume metric
$\gamma_{mn}= e_{(m}^{++}e_{n)}^{--}$ is taken into account.

If $r=0$, Eq. (\ref{B=}) does no longer follow from (\ref{YdY=}) as
$B$ does not appear in (\ref{YdY=}). Then, although the Lagrange
multiplier $\Xi$ is still present in the dynamical equations
(\ref{ddY=XiY}), their contractions with
$\overline{\Upsilon}_\Sigma$ and ${\Upsilon}^\Sigma$ cannot be used
to express $\Xi$ in terms of the covariant derivatives of
supertwistors, like in (\ref{Xi=}), but it rather produces
\begin{eqnarray}\label{DDY=YdYdY}
&& \overline{\Upsilon}_\Sigma \nabla * \nabla {\Upsilon}^\Sigma = 0
\; , \qquad
 \nabla * \nabla
\overline{\Upsilon}_\Sigma \; {\Upsilon}^\Sigma = 0 \; .
\end{eqnarray}
In the light of Eq. (\ref{YdY=}), Eqs. (\ref{DDY=YdYdY}) imply
\begin{eqnarray}\label{DYDY=0}
 \nabla \overline{{\Upsilon}}_\Sigma \wedge * \nabla
{\Upsilon}^\Sigma = 0 \quad \text{for} \quad r=0\; .
\end{eqnarray}

The existence of the invariant integral form (\ref{Om3-4}) makes
the $N=4$ supertwistor space $\mathbb{CP}^{(3|4)}$ a Calabi-Yau
supermanifold; this is needed to relate the sigma model to the
topological `B-model' \cite{Witten03}. To reproduce the MHV
amplitudes of the Yang-Mills theory the twistor string model based
on  Eq. (\ref{TWS-W}) has to be enriched by D-instanton
contributions \cite{Witten03} (that the Berkovits model seeks to
avoid). We will not need nor consider these details below.

\setcounter{equation}0
\section{$D=4$ $N=4$ superspace formulation of the supertwistor string}\label{SecIII}

\subsection{Siegel's closed supertwistor string model as a model in D=4 N=4
superspace}\label{Subsect.Siegel.Superspace}

We show here that the (non Yang-Mills part of) Siegel's closed
twistor string, $e^{++}\wedge
\bar\Upsilon_\Sigma\nabla\Upsilon^\Sigma$ in Eq. (\ref{TWS-S}),
has a transparent $D=4$, $N=4$ superspace form,
\begin{eqnarray}\label{S-0S}
S_S &=& \int_{W^2} \left[ e^{++} \wedge
\hat{\Pi}^{\dot{\alpha}\alpha}
 \bar{\lambda}_{\dot{\alpha}}\lambda_{\alpha} + d^2\xi L_{G}\right] \; ,
\end{eqnarray}
where $\hat{\Pi}^{\dot{\alpha}\alpha}\equiv d\xi^m
\Pi_m^{\dot{\alpha}\alpha} \equiv d\tau
\Pi_\tau^{\dot{\alpha}\alpha} + d\sigma
\Pi_\sigma^{\dot{\alpha}\alpha}$ is the pull-back to the
worldsheet $W^2$ of the flat supervielbein on  $D=4$, $N=4$
superspace,
\begin{eqnarray}
\label{Pi4D} \Pi^{\dot{\alpha}\alpha} &:= & dx^{\dot{\alpha}\alpha} - i
d\theta_i^{\alpha} \bar{\theta}^{\dot{\alpha}i} + i \theta_i^{\alpha}
d\bar{\theta}^{\dot{\alpha}i} \; , \qquad
\end{eqnarray}
where $i=1,2,3,4$. $\Pi^{\dot\alpha\alpha}$ can also be written in
terms of a left or right chiral coordinate basis,
\begin{eqnarray}\label{Pi4DXLR}
& \Pi^{\dot{\alpha}\alpha} := & dx_L^{\dot{\alpha}\alpha} - 2i d\theta_i^{\alpha}
\bar{\theta}^{\dot{\alpha}i} = dx_R^{\alpha\dot{\alpha}} + 2i \theta_i^{\alpha}
d\bar{\theta}^{\dot{\alpha}i} \; , \qquad  \\
\label{XL=} && x_L^{\dot{\alpha}\alpha} :=  x^{\dot{\alpha}\alpha} +
i \theta_i^{\alpha} \bar{\theta}^{\dot{\alpha}i} =
(x_R^{\dot{\alpha}\alpha})^\ast \;,\\
\label{XR=} &&x_R^{\dot{\alpha}\alpha} :=  x^{\dot{\alpha}\alpha}
-i \theta_i^{\alpha} \bar{\theta}^{\dot{\alpha}i} =
(x_L^{\dot{\alpha}\alpha})^\ast \; .
\end{eqnarray}
The bosonic spinors $\lambda_\alpha,\bar\lambda_{\dot\alpha}$ in
(\ref{S-0S}) are auxiliary. Their equations of motion
\begin{equation}\label{lambdaeq}
e^{++}\wedge \Pi^{\dot\alpha\alpha} \lambda_\alpha=0\quad \text{or}
\quad \Pi^{\dot\alpha\alpha}_{--}\lambda_\alpha=0
\end{equation}
are non-dynamical and imply
\begin{equation}\label{auxi}
\Pi^{\dot\alpha\alpha}_{--}\sim\bar\lambda^{\dot\alpha}\lambda^\alpha
\quad,
\end{equation}
which solves the Virasoro constraint
\begin{equation}\label{Vconst}
\Pi^{\dot\alpha\alpha}_{--}\Pi_{--\dot\alpha\alpha}=0 \; .
\end{equation}

It may be easily checked that the action (\ref{S-0S}) is equivalent
to (\ref{TWS-S}). Indeed, by using Leibniz rule ($
dx^{\dot{\alpha}\alpha}\, \lambda_{\alpha}= d(x^{\dot{\alpha}\alpha}
\lambda_{\alpha}) - x^{\dot{\alpha}\alpha} d\lambda_{\alpha}$, etc.)
Eq. (\ref{S-0S}) can be written in the form
\begin{eqnarray}\label{TWS-Sc}
S &=& \int e^{++} \wedge \left(d{\mu}^{\dot{\alpha}}\,
\bar{\lambda}_{\dot{\alpha}} - d\lambda_\alpha \, \bar{\mu}^\alpha
- 2i
d\eta_i \, \bar{\eta}^i \right) + d^2\xi L_{G} \nonumber \\
 &=&
 \int e^{++} \wedge
d(\Upsilon^\Sigma ) \, \Bar{\Upsilon}_\Sigma
  + d^2\xi L_{G} \; , \qquad
\end{eqnarray}
where  the components of the supertwistor are related to the
superspace coordinates by the following supersymmetric
generalization \cite{FS78-83} of the Penrose incidence relation
\cite{Pen}
\begin{eqnarray}\label{Y=XY}
\mu^{\dot{\alpha}} = x_L^{\dot{\alpha}\alpha} \lambda_{\alpha}:=
(x^{\dot{\alpha}\alpha} + i
\theta_i^{\alpha}\bar{\theta}^{\dot{\alpha}i} ) \lambda_{\alpha}\; ,
\quad \eta_i = \theta_i^{\alpha} \lambda_{\alpha}\; .  \quad
\end{eqnarray}
Eqs. (\ref{Y=XY}) give the general solution of the constraint
$\Bar{\Upsilon}_\Sigma {\Upsilon}^\Sigma = 0$, which allows us to
use Eq. (\ref{YY=0}) instead of (\ref{Y=XY}).

\subsection{$\kappa$-symmetry}\label{subsect.kappa-symmetry}

The action (\ref{S-0S}), involving the superspace coordinate
fields and the auxiliary bosonic spinor fields
$\lambda(\xi),\overline{\lambda}(\xi)$, is thus equivalent to the
twistor action (\ref{TWS-Sc}). Ignoring the Yang-Mills current
variables in $L_{G}$ and the auxiliary one-form $e^{++}$, one sees
that the superspace action (\ref{S-0S}) contains, besides the
auxiliary $e^{++}$, {\bf 8} (4+4) real bosonic and {\bf 16}
($4\times 4$) real fermionic variables, while the twistor one
(\ref{TWS-Sc}) contains instead ${\bf 8}$ bosonic plus {\bf 8}
($4\times 2$) fermionic supertwistor variables subject to one
bosonic constraint, Eq. (\ref{YY=0}). This mismatch indicates the
presence of one bosonic and eight fermionic gauge symmetries in
the superspace action (\ref{S-0S}).

The action is invariant under reparametrization as well as under
the scaling (\ref{Weyl}), $\lambda'=\lambda e^{\beta'}$ (see Eq.
(\ref{scaling})). Besides, there is a bosonic gauge symmetry under
the $U(1)$ phase transformations  of the spinor field
$\lambda_\alpha$. This is the same gauge symmetry possessed by the
supertwistor action (\ref{TWS-Sc}), there generated by the first
class constraint (\ref{YY=0}). Let us now show that the superspace
action indeed possesses an {\bf 8}-parametric fermionic
$\kappa$-symmetry.

Varying the action (\ref{S-0S}) we find (mainly ignoring the
Yang-Mills current part $L_{G}$ which does not depend neither on
supertwistors nor on the superspace coordinate functions)
\begin{eqnarray}\label{vS-0S}
& \delta S  = \int\limits_{W^2} \delta e^{++} \wedge
\left(\hat{\Pi}^{\dot{\alpha}\alpha}
\bar{\lambda}_{\dot{\alpha}}\lambda_{\alpha} +
{\delta (d^2\xi L_{G}) \over \delta e^{++}} \right) - \nonumber \\
&- \int\limits_{W^2}  d(e^{++} \lambda_{\alpha}
\bar{\lambda}_{\dot{\alpha}} )\;
 \; (\delta\hat{x}^{\dot{\alpha}\alpha} - i
\delta\hat{\theta}_i^{\alpha} \hat{\bar{\theta}}{}^{\dot{\alpha}i} +
i \hat{\theta}_i^{\alpha} \delta {}
\hat{\bar{\theta}}{}^{\dot{\alpha}i} ) \qquad
 \nonumber \\
 &+\int\limits_{W^2}
 e^{++} \wedge
\hat{\Pi}^{\dot{\alpha}\alpha} ( \bar{\lambda}_{\dot{\alpha}}\delta
\lambda_{\alpha} + \delta
\bar{\lambda}_{\dot{\alpha}}\lambda_{\alpha} ) \qquad
 \nonumber \\ &-2i  \int\limits_{W^2} e^{++} \wedge \left( d\theta^\alpha_i
 \lambda_{\alpha} \; \delta\bar{\theta}^{\dot{\alpha}i}\,
 \bar{\lambda}_{\dot{\alpha}} +
d\bar{\theta}^{\dot{\alpha}i}\,
 \bar{\lambda}_{\dot{\alpha}}
\; \delta \theta^\alpha_i
 \lambda_{\alpha} \right) \; . \qquad
\end{eqnarray}
The fact that only {\bf 8} ($\delta \theta^\alpha_i
\lambda_{\alpha}$ and its c.c.)  out of the 8+8 independent
fermionic variations ($\delta\theta^\alpha_i\; \text{and} \;
\delta\bar\theta^{\dot\alpha i}= (\delta\theta^\alpha_i)^\ast $,
$i=1,2,3,4$) enter {\it effectively} in the action variation
(\ref{vS-0S}) shows that the action (\ref{S-0S}) possess eight local
fermionic $\kappa$-symmetries. Explicitly they read
\begin{eqnarray}\label{kappaS}
\delta_\kappa x^{\dot{\alpha}\alpha}= i \delta_\kappa
\theta^\alpha_i \, \bar{\theta}^{\dot{\alpha}i} - i \theta^\alpha_i
 \delta_\kappa \bar{\theta}^{\dot{\alpha}i}  \; , \qquad \nonumber
 \\
\delta_\kappa \theta^\alpha_i \,= \kappa_i \lambda^{\alpha}\; ,
\qquad  \delta_\kappa \bar{\theta}^{\dot{\alpha}i} \, =
\bar{\kappa}^{i} \bar{\lambda}^{\dot{\alpha}} \; , \\ \nonumber
\delta_\kappa \lambda^{\alpha}= \delta_\kappa
\bar{\lambda}^{\dot{\alpha}}= \delta_\kappa e^{++}=0 \; . \qquad
\end{eqnarray}
The $\kappa$-symmetry transformations in the form (\ref{kappaS})
are clearly irreducible (see \cite{STV89} for their interpretation
as worldline supersymmetry) in contrast with the standard
$\kappa$-symmetry \cite{JdA+Lu82,Sie83} with
\begin{eqnarray}\label{kappaSS}
&& \delta_\kappa \theta^\alpha_i \, = \kappa_{\dot{\alpha}i}
\Pi_{\!_{--}}^{\dot{\alpha}\alpha} \quad,\quad
\delta_\kappa\bar\theta^{\dot\alpha
i}=\Pi_{--}^{\dot\alpha\alpha}\bar\kappa_{\alpha}^{i} \\
\nonumber && (\Pi_{\!_{--}}^{\dot{\alpha}\alpha} :=
 \nabla_{\!\!_{--}} x^{\dot{\alpha}\alpha} - i \nabla_{\!\!_{--}} {\theta}^{\alpha}\,
 \bar{\theta}^{\dot{\alpha}\alpha} + i {\theta}^{\alpha}
 \nabla_{\!\!_{--}} \bar{\theta}^{\dot{\alpha}\alpha})\; .
\end{eqnarray}

Clearly, the irreducible transformations (\ref{kappaS}) can be
obtained from the standard $\kappa$-symmetry in its first order
form by substituting
$\bar{\lambda}{}^{\dot{\alpha}}\lambda^\alpha$ for
$\Pi_{\!_{--}}^{\dot{\alpha}\alpha}$ with
$\kappa_i=\kappa_{i\dot\alpha}\bar\lambda^{\dot\alpha}$. Equation
$\Pi_{\!_{--}}^{ \dot{\alpha}\alpha} \propto
\bar{\lambda}^{\dot{\alpha}}\lambda^\alpha$ indeed holds on the
mass shell for the dynamical system (\ref{S-0S}), see Eq.
(\ref{auxi}).

\subsection{The supertwistor string as a formulation of the
tensionless string}

The fact that the action (\ref{TWS-Sc}) corresponds to a
tensionless superstring  was noticed in \cite{Siegel04}. We have
seen above that the twistor string action (\ref{TWS-Sc}) is
equivalent to the superspace action (\ref{S-0S}) which includes
the bosonic spinors $\lambda_\alpha$ as auxiliary variables. Our
next observation  is that the action (\ref{S-0S}) is simply the
null-superstring action of \cite{BZnull,BZnullr}. Indeed, although
that action was written in terms of $D=4$ Lorentz harmonic
variables $(v_\alpha^- , v^+_\alpha)\in
SL(2,\mathbf{C})=Spin(1,3)$, and (\ref{S-0S}) contains instead one
bosonic spinor $\lambda_\alpha$, the second harmonic $v^+_\alpha$
was not involved in the null-superstring action of
\cite{BZnull,BZnullr}. Furthermore, the only constraint that is
imposed on these $D=4$ spinorial harmonics \cite{B90} is
\begin{eqnarray}\label{vv=1} v^{\alpha
-}v_{\alpha}^+=1\; .
\end{eqnarray}
If this is considered as a condition on $v^+_\alpha$, then
$v_\alpha^-$ is just an arbitrary but nonvanishing bosonic spinor
and can be identified with $\lambda_\alpha$. Then, with
$\lambda_\alpha =v_\alpha^-$, and defining a worldsheet density
$\rho^{++m}$ by
 $e^{++} \wedge d:=
d^2\xi \, \rho^{++m}
\partial_m$, one can write the action (\ref{S-0S}) in
the form
\begin{eqnarray}\label{S(BZ)}
S =  &\int\limits_{W^2} d^2\xi \, \rho^{++m}
\Pi^{\dot{\alpha}\alpha}_m \bar{v}_{\dot{\alpha}}^- v_{\alpha}^-
\equiv \int\limits_{W^2} e^{++} \wedge \Pi^{\dot{\alpha}\alpha}
\bar{v}_{\dot{\alpha}}^- v_{\alpha}^- \; , \qquad
 \end{eqnarray}
which is exactly the `twistor-like' tensionless superstring action
in \cite{BZnull,BZnullr}.

\subsection{Berkovits supertwistor open string model and the open
tensionless superstring in an enlarged superspace}

The above observations indicate that Berkovits open string version
of the twistor string model should correspond to the open
null-string. We show now that this open tensionless superstring
can be defined on the direct product of two $N=4$ superspaces
\footnote{Notice that the open tensionless string in a generalized
$D=4$, $N=1$ superspace enlarged by the tensorial central charge
coordinates $y^{mn}= - y^{nm}$ was considered in \cite{Zh+Bg}.}.

Let us consider the left and right moving supertwistors already
restricted by the constraints
\begin{eqnarray}\label{YY=2+}
\Bar{\Upsilon}^-_{\Sigma }\Upsilon^{-\Sigma } \equiv
\bar{\lambda}^-_{\dot{\alpha}} \mu^{-\dot{\alpha}}
 - \bar{\mu}^{-{\alpha}}
{\lambda}^-_{{\alpha}} - 2i \eta^-_i\, \bar{\eta}^{-i} = 0 \; , \qquad \\
 \label{YY=2-}
 \Bar{\Upsilon}^+_{\Sigma }\Upsilon^{+\Sigma }
 \equiv \bar{\lambda}^+_{\dot{\alpha}} \mu^{+\dot{\alpha}}
 - \bar{\mu}^{+{\alpha}}
{\lambda}^+_{{\alpha}} - 2i \eta^+_i\, \bar{\eta}^{+i} = 0 \; .
\qquad
\end{eqnarray}
Then the connection ${B}=e^{++}{B}_{++}+ e^{--}{B}_{--}$ (which in the original
formulation of Eq. (\ref{TWS-B}) \cite{NB04} reproduces the above two constraints as
the Euler-Lagrange equations for $B_{--}$ and $B_{++}$) disappears from the action,
which takes the form
\begin{eqnarray}\label{TWS-B+}
 S_{B} &=& \int_{W^2}\left[ e^{++} \wedge \Bar{\Upsilon}^-_{\Sigma } d\Upsilon^{-\Sigma}
 - e^{--} \wedge \Bar{\Upsilon}^+_{\Sigma}
 d\Upsilon^{+\Sigma}\right]
   + \quad \nonumber \\
 && + \int_{W^2} d^2\xi L_{G} \; .
\end{eqnarray}
The local Lorentz symmetry $SO(1,1)$ of the action (\ref{TWS-B+})
as well as its two $U(1)$ gauge symmetries, one acting on
$\Upsilon^{+\Sigma}$ and $\Bar{\Upsilon}^+_{\Sigma}$ and the other
on $\Upsilon^{-\Sigma}$ and $\Bar{\Upsilon}^-_{\Sigma}$, hold true
due to the constraints (\ref{YY=2+}), (\ref{YY=2-}) imposed on
$\Upsilon^-$ and $\Upsilon^+$. The action also has an overall
scaling gauge symmetry under $\Upsilon^{+,-} \mapsto
e^{\gamma}\Upsilon^{+,-}$, $e^{\pm\pm} \mapsto e^{-2\gamma}
e^{\pm\pm}$.

Now, following with this action the same steps as in Sec.
\ref{Subsect.Siegel.Superspace} , but in reverse order, we recover
a counterpart of (\ref{S-0S}) for the open twistor string action
of Eq. (\ref{TWS-B}). Starting from Eq. (\ref{TWS-B+}) and solving
the constraints (\ref{YY=2+}), (\ref{YY=2-}) by
\begin{eqnarray} \label{Y=XY-}
\mu^{- \dot{\alpha}} &=& x_{(l)L}^{\dot{\alpha}\alpha}
\lambda^-_{\alpha}:= (x_{(l)}^{\dot{\alpha}\alpha} + i
\theta_{(l)}{}_i^{\alpha}\bar{\theta}_{(l)}^{\dot{\alpha}i} )
\lambda^-_{\alpha}\; , \quad \nonumber \\  \eta^-_{i} &=&
\theta_{(l)}{}_i^{\alpha} \lambda^-_{\alpha}\; ; \quad \\
\label{Y=XY+} \mu^{+ \dot{\alpha}} &=& x_{(r)L}^{\dot{\alpha}\alpha}
\lambda^+_{\alpha}:= (x_{(r)}^{\dot{\alpha}\alpha} + i
\theta_{(r)}{}_i^{\alpha}\bar{\theta}_{(r)}^{\dot{\alpha}i} )
\lambda^+_{\alpha}\; , \quad  \quad \nonumber \\  \eta^+_{i} &=&
\theta_{(r)}{}_i^{\alpha} \lambda^+_{\alpha}\; , \quad
\end{eqnarray}
one finds that the action $S_B$ of (\ref{TWS-B+}) is equivalent to
\begin{eqnarray}\label{S-0B}
S &=& \int_{W^2}  (e^{++} \wedge \hat{\Pi}_{(l)}^{\dot{\alpha}\alpha}
 \bar{\lambda}^{-}_{\dot{\alpha}}\lambda^{-}_{\alpha} - e^{--} \wedge
\hat{\Pi}_{(r)}^{\dot{\alpha}\alpha}
\bar{\lambda}^{+}_{\dot{\alpha}}\lambda^{+}_{\alpha}) \quad
\nonumber \\ &&
 + \int_{W^2} d^2\xi L_{G} \; ,
\end{eqnarray}
where
\begin{eqnarray}
\label{Pi(l)} \Pi_{(l)}^{\dot{\alpha}\alpha} &:=& dx_{(l)}^{\dot{\alpha}\alpha} - i
d\theta_{(l)}{}_i^{\alpha} \bar{\theta}_{(l)}^{\dot{\alpha}i} + i
\theta_{(l)}{}_i^{\alpha} d\bar{\theta}_{(l)}^{\dot{\alpha}i} \; , \quad\nonumber
\\ \label{Pi(r)}
\Pi_{(r)}^{\dot{\alpha}\alpha} &:=& dx_{(r)}^{\dot{\alpha}\alpha} - i
d\theta_{(r)}{}_i^{\alpha} \bar{\theta}_{(r)}^{\dot{\alpha}i} + i
\theta_{(r)}{}_i^{\alpha} d\bar{\theta}_{(r)}^{\dot{\alpha}i} \; . \quad
\end{eqnarray}
The action (\ref{S-0B}) contains two sets of coordinate functions,
$x_{(l)}^{\alpha\dot{\alpha}}, \theta_{(l)}{}_i^{\alpha},
\bar{\theta}_{(l)}^{\dot{\alpha}i}$ and
$x_{(r)}^{\alpha\dot{\alpha}}, \theta_{(r)}{}_i^{\alpha},
\bar{\theta}_{(r)}^{\dot{\alpha}i}$, corresponding to two copies
of $D=4$, $N=4$ superspace. Looking at the dynamics implied by
(\ref{S-0B}) it is seen that one set
($x_{(l)}^{\alpha\dot{\alpha}}, \theta_{(l)}{}_i^{\alpha},
\bar{\theta}_{(l)}^{\dot{\alpha}i}$) contains the left- and the
other ($x_{(r)}^{\alpha\dot{\alpha}}, \theta_{(r)}{}_i^{\alpha},
\bar{\theta}_{(r)}^{\dot{\alpha}i}$) the right-moving fields, as
indicated by the subindexes $l,r$.

We note in passing  that such a double set of variables, but for
$N=1$, was used in \cite{Ivanov+Isaev88} to write an equivalent form
of the $N=2$ Green-Schwarz superstring action.

In the above discussion on the open twistor string, the doubling of
the superspace variables seems to play an auxiliary r\^ole  as far
as the YM vertex operators are associated with the boundary of the
open string. The two sets of $D=4$ $N=4$ superspace coordinate
functions are needed to formulate the action in its spacetime form.
The boundary conditions (\ref{TWS-Bbc}) identify these coordinate
functions modulo the (two copies of the) $\kappa$-symmetry,
\begin{eqnarray}\label{kappaSL}
& \delta_\kappa x_{(l)}^{\dot{\alpha}\alpha}= i \delta_\kappa \theta_{(l)}{}^\alpha_i
\, \bar{\theta}_{(l)}^{\dot{\alpha}i} - i \theta_{(l)}{}^\alpha_i
 \delta_\kappa \bar{\theta}_{(l)}^{\dot{\alpha}i}  \; , \qquad \nonumber
 \\
& \delta_\kappa \theta_{(l)}{}^\alpha_i \,= \kappa^{+}_i \lambda^{-\alpha}\; , \qquad
\delta_\kappa \bar{\theta}_{(l)}^{\dot{\alpha}i} \, = \bar{\kappa}^{+i}
\bar{\lambda}^{-\dot{\alpha}} \; ,
\\ \label{kappaSR} & \delta_\kappa x_{(r)}^{\dot{\alpha}\alpha}= i
\delta_\kappa \theta_{(r)}{}^\alpha_i \, \bar{\theta}_{(r)}^{\dot{\alpha}i} - i
\theta_{(r)}{}^\alpha_i
 \delta_\kappa \bar{\theta}_{(r)}^{\dot{\alpha}i}  \; , \qquad \nonumber
 \\
& \delta_\kappa \theta_{(r)}{}^\alpha_i \,= \kappa^{-}_i \lambda^{+\alpha}\; , \qquad
\delta_\kappa \bar{\theta}_{(r)}^{\dot{\alpha}i} \, = \bar{\kappa}^{-i}
\bar{\lambda}^{+\dot{\alpha}} \; , \\
& \nonumber \delta_\kappa \lambda^{\pm\alpha}= \delta_\kappa
\bar{\lambda}^{\pm\dot{\alpha}}= \delta_\kappa e^{\pm\pm}=0 \; ,
\qquad
\end{eqnarray}
and reparametrization symmetry transformations characteristic of the
action (\ref{S-0B}).

\setcounter{equation}0
\section{On a possible parent tensionful  superstring action for the twistor
string}\label{Sect.Parent.tensionful} The null-superstring mass
spectrum is known to be continuous \cite{BZnullb,BZnullr}. To obtain
a discrete spectrum, one should rather quantize the tensionless
limit of a tensionful superstring. These two zero tension
superstrings are different in the set of variables used to build the
quantum theory. The null-superstring \cite{BZnullb,BZnullr} is
quantized in terms of particle-like variables, momentum and
coordinates, while the quantum theory of the tensionless limit of a
superstring (often called just `tensionless superstring') is
formulated in terms of stringy oscillators
\cite{Lindstrom03,Bonelli,SagnottiTsulaia}.

The calculations of the tree YM diagrams from the twistor string
models \cite{Witten03,NB04,Siegel04}, in particular the choice of
the vertex operators and the discussions on contributions to
conformal anomaly, clearly use stringy variables rather than  the
particle-like null-superstring ones and, thus, deal with a
tensionless limit of some tensionful superstring rather than with
the intrinsically tensionless superstring {\it i.e.}
null-superstring.

Thus it is natural to ask: which is the tensionful superstring
action the tensionless limit of which leads to the twistor string
one? Such a problem was posed by Siegel \cite{Siegel04}, who
proposed the tensionful QCD string \cite{Siegel-QCD} as the
bosonic part of such a parent superstring; fermions were not
considered. In our present perspective, the above problem
corresponds to looking for the tensionful parent of the $N=4$
version of the $N=1$ tensionless superstring action of
\cite{BZnull,BZnullr}, Eq. (\ref{S(BZ)}).

\subsection{ From tensionful D=4 superstrings to N=1,2
counterparts of the supertwistor string}

The $N$=1 and $N$=2 versions of the null-superstring superspace
action (\ref{S(BZ)}) can be obtained as tensionless limits of the
action of the Lorentz harmonics formulation \cite{BZ91} of the
$N=1$ and $N=2$ $D=4$ Green-Schwarz superstrings
\begin{widetext}
\begin{eqnarray}\label{S(BZst)}
S= {1\over 4 \pi \alpha^\prime} \int_{W^2} [e^{++} \wedge
\Pi^{\dot{\alpha}\alpha}  \bar{v}_{\dot{\alpha}}^-v_{\alpha}^- -
e^{--} \wedge \Pi^{\dot{\alpha}\alpha}
\bar{v}_{\dot{\alpha}}^+v_{\alpha}^+ - e^{++} \wedge  e^{--}] -
{1\over 4\pi \alpha^\prime} \int_{W^2} \widehat{B}_2\; ,
 \end{eqnarray}
\end{widetext}
where the bosonic spinors $v_{\alpha}^-$ and $v_{\alpha}^+$ are
relatively normalized by the `harmonicity conditions'
(\ref{vv=1}), \mbox{$v^{\alpha -}v_{\alpha}^+=1$}, and the last
contribution in (\ref{S(BZst)}) is the Wess-Zumino term. This is
defined by the pull-back to $W^2$ of the two-form gauge potential
$B_2$ on flat superspace and that provides the superspace
generalization of the NS-NS or Kalb-Ramond field. This two-form
obeys the constraints \footnote{The expression of $H_3$ shows that
it is a Chevalley-Eilenberg (CE) three-cocycle for the superspace
algebra cohomology \cite{JdeA+PT89}.}
\begin{eqnarray}\label{H3(N2)}
& H_3= dB_2= -2i \Pi^a \wedge (d\theta^1 \wedge \sigma_a
d\bar{\theta}^1 - d\theta^1 \wedge \sigma_a d\bar{\theta}^2) \quad
\nonumber \\ & \qquad  \text{for} \; N=2 \; , \qquad\\
\label{H3(N1)}
& H_3= dB_2= -2i \Pi^a \wedge d\theta \wedge \sigma_a d\bar{\theta}
\quad \text{for} \; N=1 \; . \quad
\end{eqnarray}

The $N=1,2$ versions of the null-superstring action (\ref{S(BZ)})
can be obtained from (\ref{S(BZst)}) by taking the tensionless
limit $ \alpha^\prime\mapsto \infty $ {\it while} keeping
$\frac{e^{++}}{\alpha^\prime}$ finite. Thus, before setting
$\alpha^\prime\mapsto \infty $ we redefine $e^{++}\rightarrow 4\pi
\alpha^\prime e^{++}$, $e^{--}\rightarrow e^{--}/(4\pi
\alpha^\prime)$. In this way, taking the tensionless limit
$\alpha^\prime\mapsto \infty$ one finds that the Wess-Zumino term
and the `cosmological' $e^{++}\wedge e^{--}$ term vanish as
$1/\alpha^\prime\mapsto 0 $. Similarly, the second term in
(\ref{S(BZst)}) also goes to zero as $1/(\alpha^\prime)^2 \mapsto
0 $, while after the redefinition the first term becomes
$\alpha^\prime$ independent and produces the tensionless
superstring action (\ref{S(BZ)}).

The problem with the $N$=4 tensionless superstring, which is
equivalent to the twistor string model
\cite{Witten03,NB04,Siegel04}, is that the corresponding $N=4$
tensionful superstring, which would be the counterpart of the
$N=1,2$ actions (\ref{S(BZst)}) possessing a $2N$ parametric
$\kappa$-symmetry, is not known. This problem may be traced to the
mismatch between the on-shell bosonic and fermionic degrees of
freedom of such a hypothetical $N=4$ superstring constructed from
$\hat{x}^a$, $\hat{\theta}^{\alpha 1}$, $\ldots$,
$\hat{\theta}^{\alpha 4}$ and their complex conjugates ($4-2= 2$
bosonic and $1/2 (4\times 2)=4$ fermionic degrees of freedom).
Geometrically, the problem is reflected by the absence of the
$D=4, N=4$ counterpart of the CE three-cocycles $H_3=dB_2$
\cite{JdeA+PT89} that do exist in $D=4, N=1,2$ superspaces. Such a
closed three form would be needed to construct the Wess-Zumino
term, a necessary ingredient of a $\kappa$-symmetric {\it
tensionful} superstring action in the superspace of the usual type
(see \cite{30/32} for a superstring action without Wess-Zumino
term in an enlarged tensorial superspace and \cite{CAIPB2000} for
a discussion of WZ terms and extended superspaces).

\subsection{ From  D=10 N=1 superstring to the supertwistor
string}\label{Subsect.From superstring}

Such a three-cocycle does exist for the $D=10$ $N=1$ supersymmetry
algebra, allowing for the existence of the heterotic superstring
\cite{Heterotic}. It is given by
\begin{widetext}
\begin{eqnarray}\label{H3(10D)}
&  D=10\; , \;  N=1 \; : \qquad H_3=dB_2 = -2i \Pi^{\underline{a}}
\wedge d\Theta \wedge \Sigma_{\underline{a}} d\Theta\; . \quad
\end{eqnarray}
The $N=1$ $D=10$ superstring contains (as the $D$=4, $N$=4 one) 16
fermionic Majorana-Weyl coordinate functions
\begin{eqnarray}\label{Th=}
\Theta^{\underline{\alpha}} = \left( \begin{matrix}
\theta_i^{{\alpha}}  \cr \bar{\theta}^{i}_{\dot{\alpha}}
\end{matrix}\right)\equiv  \left( \begin{matrix} \theta_i^{{\alpha}} \cr ( \theta_{\alpha\, i})^*
\end{matrix}\right) \; , \quad  \underline{\alpha}= 1, \ldots , 16 \; , \quad
{\alpha}= 1, 2 \;  , \quad {\dot{\alpha}} = 1, 2 \; , \;  \quad
\quad i=1, \ldots 4\; ,
 \end{eqnarray}
and ten bosonic coordinate functions $X^{\underline{a}}$ which can
be split as
\begin{eqnarray}\label{X=x,X}
X^{\underline{a}}= (x^a \; ,\; X^I)\; , \qquad \underline{a}= 0,1 \ldots , 10 \; ,
\quad a=0,1,2,3 \; , \quad I=1, \ldots , 6  \; .
\end{eqnarray}
The $D=10$ ($16\times 16$) sigma-matrices can be chosen in the
form
\begin{eqnarray}\label{10DSigma}
& \Sigma_{\alpha\beta}^{\underline{a}} =
(\Sigma_{\alpha\beta}^{{a}}\; , \Sigma_{\alpha\beta}^{I}) \; , \quad
\Sigma_{\alpha\beta}^{{a}}= \left(\begin{matrix} 0 &
\sigma_{\alpha\dot{\beta}}^{{a}}\delta^i_j \cr
\sigma^{{a}{\dot{\alpha}}{\beta}}\delta^j_i & 0
\end{matrix}\right)
 \; , \quad  \Sigma_{\alpha\beta}^{{I}}=
\left(\begin{matrix}  \epsilon_{\alpha\beta} \tilde{\rho}^{I ij} & 0 \cr
 0 & - \epsilon_{\dot{\alpha}\dot{\beta}} {\rho}^{I}_{ ij}
\end{matrix}\right)\; ,
 \end{eqnarray}
where ${\rho}^{I}_{ ij}$ and $\tilde{\rho}^{I ij}$ are the $SO(6)$
Clebsch-Gordan coefficients ${\rho}^{I}_{ ij}\tilde{\rho}^{I
i^\prime j^\prime}= - 4 \delta_i^{[i^\prime} \delta_j^{\,
j^\prime]}$ (see {\it e.g.} \cite{GSW}).

The Lorentz harmonics formulation of  the $D=10, N=1$ superstring is
characterized by the action \cite{BZ91,BZ94} which can be written in
the form
\begin{eqnarray}\label{S(BZst)10}
S= {1\over 4 \pi \alpha^\prime} \int_{W^2} [e^{++} \wedge  \Pi^{\underline{a}}
u_{\underline{a}}{}^{--} -  e^{--} \wedge
 \Pi^{\underline{a}}
u_{\underline{a}}{}^{++} -  e^{++} \wedge e^{--}] - {1\over 4\pi \alpha^\prime}
\int_{W^2} \hat{B}_2\;  \qquad
 \end{eqnarray}
involving the worldvolume fields in the pull-back of the NS-NS two
form $B_2$ (\ref{H3(10D)}) to $W^2$ and two auxiliary lightlike
vector fields, $u^{++}_{\underline{a}}$, $u_{\underline{a}}^{--}$,
the counterparts of $ v_{\alpha}^- \bar{v}_{\dot{\alpha}}^-$ and $
v_{\alpha}^+ \bar{v}_{\dot{\alpha}}^+$ in $D$=4 (\ref{S(BZst)})
\footnote{One could also write another twistor-like action for
$D=10$ superstrings by using two unconstrained bosonic spinors
\cite{SP92}. In our notation, it reads $ {1\over 4 \pi \alpha'}
\int_{W^2} [e^{++} \wedge \Pi^a (\lambda^-\Sigma_a\lambda^-) -
e^{--} \wedge
 \Pi^a (\lambda^+\Sigma_a\lambda^+) -  e^{++} \wedge e^{--}
(\lambda^-\Sigma_a\lambda^-)(\lambda^+\Sigma^a\lambda^+)] -
{1\over 4\pi \alpha^\prime} \int_{W^2} \hat{B}_2\; $. Then, and in
contrast with (\ref{S(BZst)10}), the $\kappa$-symmetry of that
action would not be irreducible.}. These `vector Lorentz
harmonics' \cite{Sok} may be considered as composites of the
$D=10$ spinorial harmonics or spinor moving frame variables
\cite{gdsghs,BZ91,BZ94}. Here we only notice their lightlike
character and the relative normalization ({\it cf.} (\ref{vv=1}))
of the ten-vectors $u_{\underline{a}}{}^{\pm\pm}$,
\begin{eqnarray}\label{uu=}
u^{\underline{a}--}u_{\underline{a}}{}^{--}=0 \; , \qquad
u^{\underline{a}++}u_{\underline{a}}{}^{++}=0 \; , \qquad
u^{\underline{a}--}u_{\underline{a}}{}^{++}=2 \; . \qquad
 \end{eqnarray}

Taking the $\alpha^\prime \mapsto \infty$ limit in Eq.
(\ref{S(BZst)10}) after the $e^{++}\rightarrow 4\pi \alpha^\prime
e^{++}$, $e^{--}\rightarrow e^{--}/(4\pi \alpha^\prime)$
redefinition, as for the $D=4, N=1,2$ superstring action above, we
arrive at the {\it ten-dimensional} tensionless superstring action
\begin{eqnarray}\label{S(null)10}
S= \int_{W^2} e^{\!^{++}} \wedge  \Pi^{\underline{a}}
u_{\underline{a}}^{\!^{--}} \; , \qquad
u_{\underline{a}}^{{--}}u^{\underline{a}\, --}=0 \; ,
\end{eqnarray}
\end{widetext}
which involves only $u_{\underline{a}}^{--}$, one of the two
lightlike ten-dimensional vectors (\ref{uu=}). A dimensional
reduction of such an action can be done in such a manner that the
$D=4, N=4$ null superstring appears. A formal way to achieve this
is to consider the action (\ref{S(null)10}) in a frame where the
above lightlike vector $u_{\underline{a}}^{--}$ only has
nonvanishing components in the four $D=4$ Minkowski spacetime
directions,
\begin{eqnarray}\label{u10=u4}
& u_{\underline{a}}^{--}= \delta_{\underline{a}}^{\; b} u_{b}^{--}= (u_{{a}}^{--}, 0,
\ldots , 0) \; , \quad u_{{a}}^{--}u^{a--}=0 \; , \quad \nonumber \\ & \qquad
\underline{a}= 0, \ldots , 9 \; , \qquad {a}= 0, \ldots , 3\; .
 \end{eqnarray}

\subsection{Tensorial (enlarged) superspace versus standard ten
dimensional superspace.}

The above shows that the twistor string can be obtained by taking
the tensionless limit of the $D=10$ superstring action
(\ref{S(BZst)10}) and then performing a dimensional reduction down
to $D$=4. By considering the $D=10$ Green-Schwarz superstring
action as a tensionful parent of the $D=4$ twistor string, we have
allowed ourselves to enlarge $D=4$ superspace by six additional
bosonic coordinates. However, it is not clear at present whether
this enlargement is unique, and this allows us to discuss another
possible higher-dimensional superstring parent for the twistor
string.

Indeed, even if we restrict ourselves to just six additional
bosonic coordinates as above, these do not need being the
components of the $SO(6)$ vector $X^I$ implied in the enlargement
of $D=4$ to the standard $D=10$ superspace. We may consider
instead a tensorial superspace, in which the additional six
bosonic coordinates appear as the components of an antisymmetric
tensor, $Y^{\mu\nu}= - Y^{\nu\mu}$. The proper incorporation of a
SO(6) vector into the action leads naturally to an enhancement of
the symmetry from $SO(1,3)\otimes SO(6)$ to $SO(1,9)$. This
implies an embedding of our tensionless string (classically
equivalent to the twistor string) into a manifestly $SO(1,9)$
(actually, $D=10$ super-Poincar\'e) invariant theory. Similarly,
the proper enlargement of the target superspace by the
antisymmetric tensor coordinates $Y^{\mu\nu}$, which could be
split into the symmetric spin-tensor $X^{\alpha\beta}=
X^{\beta\alpha}$ and its complex conjugate (in the case of
Minkowski signature) $X^{\dot{\alpha}\dot{\beta}}=
X^{\dot{\beta}\dot{\alpha}}= (X^{\alpha\beta})^*$, results in an
enlargement of the automorphism symmetry to $GL(4,\mathbb{R})$.
The spin-tensorial representation allows to collect all ten
coordinates in a manifestly symmetric $4\times 4$ matrix,
\begin{equation}\label{tensorial}
{\mathbb{X}}^{{\alpha'}{\beta'}} := \left(\begin{matrix}   X^{\alpha\beta} &
(X^{\dot{\beta}\alpha})^T \cr X^{\dot{\alpha}\beta} & X^{\dot{\alpha}\dot{\beta}}
\end{matrix}\right)\, ,\quad \alpha',\beta'=1,\dots,n=4
\end{equation}
({\it i.e.} $\alpha'=1,2,\dot 1,\dot 2$). Such a tensorial space
was proposed by Fronsdal \cite{Fr86} to describe higher spin
fields. A dynamical realization of such a theory was found later
\cite{BLS99}, quantizing a generalized superparticle model
\cite{BL98} which has the properties of a BPS preon \cite{BPS01}
(see \cite{30/32,BPS03,B04} for further discussion).

The above analysis suggests relating our $D=4, N=4$
null-superstring (\ref{S(BZ)}) to a string model in a $N$=4
extended tensorial superspace $\widetilde{\Sigma}^{(10|4N)}$
($=\widetilde{\Sigma}^{\left(\frac{n(n+1)}{2}|nN\right)}$ for
$n=4$, see \cite{30/32})
\begin{eqnarray}\label{S10-16} && ({\mathbb{X}}^{{\alpha'}{\beta'}} \; , \;
\Theta^{{\alpha'}i}) := (X^{\dot{\beta}\beta},  X^{\alpha\beta},
X^{\dot{\alpha}\dot{\beta}} ;\, \theta^{\alpha}_i \, , \,
\bar{\theta}^{\dot{\alpha}i})\; . \qquad
\end{eqnarray}
The $D=4, N=4$ null superstring action (\ref{S(BZ)}) providing a
spacetime reformulation of the Berkovits-Siegel twistor string
action (\ref{S-0S}) can be obtained as the $w \mapsto 0$ limit of
the action (omitting the $L_{G}$ contribution)
\begin{widetext}
\begin{eqnarray}\label{S(w)-0S}
S(w) &=& \int_{W^2} e^{++} \wedge \left( {\Pi}^{\dot{\alpha}\alpha}
\lambda_{\alpha} \bar{\lambda}_{\dot{\alpha}} + {w\over 2}\,
{\Pi}^{\alpha{\beta}} \lambda_{\alpha} {\lambda}_{{\beta}} +
{\bar{w}\over 2}\, {\Pi}^{\dot{\alpha}\dot{\beta}}
\bar{\lambda}_{\dot{\alpha}} \bar{\lambda}_{\dot{\beta}} \right) \;
,
\end{eqnarray}
\begin{eqnarray}
\Pi^{\dot\alpha\alpha}&=&dx^{\dot\alpha\alpha}-
id\theta^\alpha_i\overline{\theta}^{\dot\alpha i}+
i\theta^\alpha_i d\overline{\theta}^{\dot\alpha i}\;, \qquad
\Pi^{\alpha\beta}=dx^{\alpha\beta}- 2 i
d\theta^{(\alpha}_i{\theta}^{\beta )}_{i}\;, \qquad
\Pi^{\dot\alpha\dot\beta}=dx^{\dot\alpha\dot\beta}-
2id\overline{\theta}^{(\dot{\alpha}
i}\overline{\theta}^{\dot{\beta}) i}\; ,
\end{eqnarray}
\end{widetext}
which describes a tensionless superstring in $N=4$ extended
tensorial superspace $\widetilde{\Sigma}^{(10|4N)}$ (for $w=1$
this action was first considered in \cite{Zh+U02}).

The action (\ref{S(w)-0S}), is an extended object counterpart of
the superparticle action \cite{BLS99} in tensorial superspace. It
may be related with the tensionless limit of tensionful
superstring models in enlarged superspace (higher spin extensions
of the superstring) considered in \cite{30/32}. In particular, a
direct tensionless limit of the generalized superstring model
\cite{30/32} would lead to the $w=1$ representative of the family
(\ref{S(w)-0S}) of tensionless actions. This $w=1$ action can be
rewritten in the form
\begin{eqnarray}\label{S(w=1)}
S(w=1)&= S_{Sp(4|4)}  = {1\over 2} \int\limits_{W^2} e^{++} \wedge
{\Pi}^{{\alpha'}{\beta'}} \Lambda_{{\alpha'}}\Lambda_{{\beta'}}
  \; , \qquad \\
\nonumber  & {\Pi}^{{\alpha'}{\beta'}}=
 d\mathbb{X}^{{\alpha'}{\beta'}} - 2i d{\Theta}^{({\alpha'}|i} {\Theta}^{|{\beta'})i}\; , \qquad \\
 \nonumber & {\Theta}^{{\alpha'}i}  := ({\theta}^{{\alpha}}_{i}\, ,
\, {\bar{\theta}}{}^{\dot{\alpha}i}\, )\; , \quad  \Lambda_{\alpha'}:=
(\lambda_\alpha,\overline{\lambda}_{\dot\alpha})
\end{eqnarray}
which makes its $GL(4,\mathbb{R})$ symmetry manifest. It possesses
a hidden $OSp(N|8)$ symmetry which becomes manifest in its
orthosymplectic twistor presentation \cite{B02}
\begin{eqnarray}\label{S(w1-tw)}
 & S_{_{{\Sigma}^{(10|16)}}} =  {1\over 2} \int e^{++} \wedge
(d{\cal M}^{\alpha '} \, \Lambda_{\alpha '} - {\cal M}^{\alpha '}
d\Lambda_{\alpha '} - 2 i d\chi^i\, \chi^i )\, , \nonumber \\
 & {\cal M}^{\alpha '}= {\mathbb{X}}^{\alpha '\beta '} \Lambda_{\beta
'} - {i} {\Theta}^{{\alpha'}i} {\Theta}^{{\beta'}i}
\Lambda_{\beta '} \; , \qquad \nonumber \\
 & \chi^i = {\Theta}^{{\beta'}i} \Lambda_{\beta '}\;
\end{eqnarray}
(see \cite{BL98,BLS99} for the superparticle case and the
discussion in \cite{30/32}).

In the purely bosonic limit the simple redefinition
$\hat{X}^{{\alpha}{\beta}}\mapsto 1/{w}\,
\hat{X}^{{\alpha}{\beta}}$, $
\hat{X}^{\dot{\alpha}\dot{\beta}}\mapsto 1/\bar{w}\,
\hat{X}^{\dot{\alpha}\dot{\beta}}$ maps any $w\not= 0$ model to the
$w=1$ one. This implies that the symmetry of any of the $S(w\not=
0)$ actions (\ref{S(w)-0S}) includes the bosonic $Sp(8)$ group.
However, the presence of fermions breaks this identification and
makes the $w=1$ dynamical system (\ref{S(w=1)}) special as it
possesses $12$ local fermionic $\kappa$-symmetries while all other
$w\not= 0, 1$ models possess only $8$ $\kappa$-symmetries. Another
face of the same fact is that the $w=1$ model (\ref{S(w=1)}) may be
written in terms of $OSp(N|8)=OSp(4|8)$ real supertwistors
$(\mu^{\alpha'}, \, \lambda_{\alpha'} \, , \, \chi_i)$ with real
fermionic $\chi_i= (\chi_i)^*$ $i=1,\dots,4$, Eqs. (\ref{S(w1-tw)}),
while the $w\not=0,1$ models require $OSp(2N|8)=OSp(8|8)$
supertwistors $(\mu^{\alpha'}, \, \lambda_{\alpha'} \, , \, \eta_i)$
with complex fermionic components $\eta_i \not= (\eta_i)^*$ (see
\cite{BLS99} for the superparticle case),
\begin{widetext}
\begin{eqnarray}\label{S(w-tw)} & S(w) = {1\over 2} \int e^{++} \wedge
\left(d{\bar\mu}^{{\alpha}} \lambda_{\alpha} - d\lambda_{\alpha}
\bar{\mu}^{{\alpha}} + \bar{\lambda}_{\dot{\alpha}}
d{\mu}^{\dot{\alpha}} - {\mu}^{\dot{\alpha}}d
\bar{\lambda}_{\dot{\alpha}} - 2i d\eta_i (\bar{\eta}^i + {w}
\eta_i) + 2i ({\eta}_i + {\bar{w}} \bar{\eta}^i)d\bar{\eta}^i
\right)
 \, ,
\end{eqnarray}
where
\begin{eqnarray}\label{N2OSp} & \bar{\mu}^{{\alpha}} =
X^{\dot{\beta}\alpha}\bar{\lambda}_{\dot\beta} + {w}
X^{{\alpha}{\beta}} {\lambda}_{{\beta}} - i\theta^\alpha_i
\left(\bar{\theta}^{i }\bar{\lambda} + {w}\theta_i\lambda \right) \;
, \quad  {\mu}^{\dot{\alpha}} =
X^{\dot{\alpha}\beta}{\lambda}_{\beta} + \bar{w}
X^{\dot{\alpha}\dot{\beta}} \bar{\lambda}_{\dot{\beta}} - i
\bar{\theta}^{\dot{\alpha}
i } \left(\theta_i\lambda + {\bar{w}}\bar{\theta}^{i }\bar{\lambda} \right) \; ,\nonumber \\
& \eta_i=\theta_i\lambda= \theta^\alpha_i\lambda_\alpha\; ,  \qquad
\bar{\eta}^i= \bar\theta^i \bar\lambda=\bar{\theta}^{\dot{\alpha} i
}\bar{\lambda}_{\dot\alpha} \; . \qquad
\end{eqnarray}
\end{widetext}
The $w=0$ member of the above $S(\omega)$ family is equivalent to the twistor string
action (\ref{TWS-S}),
\begin{eqnarray}
\label{S(w=0)-0S} S(w=0)&=& S_S = \int_{W^2} e^{++} \wedge {\Pi}^{\dot{\alpha}\alpha}
\lambda_{\alpha} \bar{\lambda}_{\dot{\alpha}}  \; .
\end{eqnarray}
To see this, in addition to observing the coincidence of Eqs.
(\ref{S-0S}) and (\ref{S(w=0)-0S}), one should take into account
that for $w=0$ Eqs. (\ref{N2OSp}) become the incidence equations
(\ref{Y=XY}) and their c.c.; these, in turn, provide the general
solution of the constraints (\ref{YY=0}). For the general
$w\not=0$ element of the family of dynamical systems
(\ref{S(w-tw)}), Eqs. (\ref{N2OSp}) do not imply any constraints.
This corresponds to the fact that the $N=1$ superparticle
counterpart of the action (\ref{S(w-tw)}) describes an infinite
tower of massless fields of all possible helicities (`free higher
spin theory'). The significance of the fact that the twistor
string enters as a singular element of the one-complex-parameter
family of tensionless superstrings in tensorial superspace (if
any) is still to be understood. Let us finish by noticing that
there is an enlargement of the internal symmetry group of the
$w=0$ action (\ref{S(w-tw)}) from $SO(4)$ to $SU(4)$.

\section{Final remarks and discussion}
\setcounter{equation}0

 By using a twistor transform similar to the
one originally proposed for the superparticle \cite{FS78-83}, we
have seen that the twistor string model is classically equivalent
to a supersymmetric extended object in a $D=4$, $N=4$ superspace.
For Siegel's closed string version of the Berkovits model, this
action, having $8$ $\kappa$-symmetries and $16$ supersymmetries,
coincides with the $N=4$ extension of the tensionless superstring
action \cite{BZnull}. The Berkovits open-string action \cite{NB04}
describes a counterpart of the tensionless superstring in the
superspace isomorphic to a direct product of two copies of $D=4$,
$N=4$; the two copies of the coordinate functions turn out to be
identified on the open string boundary modulo gauge symmetries,
the set of which includes two copies of $8$-parametric
$\kappa$-symmetry, Eqs. (\ref{kappaSL}), (\ref{kappaSR}).

Null (or intrinsically tensionless) superstrings maintain their
conformal invariance by having a continuous spectrum
\cite{BZnullb,BZnull}. This implies that the prescription of
writing the gauge field amplitudes from \cite{Siegel04} assumes
dealing with the tensionless limit of some tensionful superstring
rather than with the null-superstring itself. This was actually
noticed in \cite{Siegel04} where a possible relation of the
twistor superstring with a model for a `QCD string'
\cite{Siegel-QCD} was discussed. However, the consideration of the
tensionful prototype of the twistor string in \cite{Siegel04} was
purely bosonic.

It is plausible to assume that a parent tensionful superstring
should have a smooth tensionless superstring limit in the sense
that both tensionful and tensionless superstrings should present
the same number of $\kappa$-symmetries. Of course, as it is known,
the zero tension limit is special in many respects. In particular
the huge enhancement of the global symmetry in this limit was
already noticed in \cite{Gross}. However, one may expect a
nonsingular limit in the sense of preserving the number of degrees
of freedom of the dynamical system. This seems to be the case for
the tensionless limit of the standard (Nambu-Goto or)
Green-Schwarz string, a limit believed to be described by a
massless higher spin theory (an infinite tower of massless higher
spin fields); see \cite{Bonelli,SagnottiTsulaia} for a discussion.
Let us stress, nevertheless, that the tensionless limit of the
usual Green-Schwarz superstring leading to the twistor string is
expected to be different. Such a zero tension limit could
accompanied by some other transformations of the variables in the
model. In any case, since Berkovits and Siegel's path integral
exponents contain the twistor string action and we have seen here
that this action is equivalent to the tensionless string one, our
results suggest that their prescription to obtain the $N=4$ SYM
diagrams from the twistor string \cite{NB04,Siegel04} provides a
third way of quantizing of the tensionless string, alternative to
the two of \cite{Lindstrom03,Bonelli}, \cite{BZnull,BZnullr}.

As we discussed in this paper, the $D=10$ $N=1$ superstring can be
considered as a tensionful candidate leading to the $D=4$ twistor
string upon dimensional reduction. At the present level of
understanding, the way from $D=10$ $N=1$ tensionful Green-Schwarz
superstring to the $D=4$ $N=1$ tensionless superstring action
(\ref{S(BZ)}) (equivalent, as we have shown here, to Siegel's
twistor string action (\ref{TWS-S})), consists in taking first a
tensionless limit then performing a dimensional reduction of the
$D=10$ tensionless superstring down to $D=4$. For such a
construction the standard $D=10$ $N=1$ superspace
$\Sigma^{(10|16)}$ is not  {\it a priori} a better starting point
than {\it e.g.}, $D=4$ $N=4$ tensorial superspace
$\widetilde{\Sigma}^{(10|16)}$, the ten bosonic coordinates of
which include the spacetime four-vector $x^\mu$ plus six tensorial
coordinates $y^{\mu\nu}$. These can be treated as spin degrees of
freedom \cite{Fr86,BLS99} or as conjugated
\cite{CAIPB2000,BL98,30/32} to the topological charges
 of superbranes \cite{AGIT89}.

This is a good place to discuss the possible higher dimensional
generalizations of the supertwistor string (a problem also posed
in \cite{Bars04} in the context of two-time physics). The
generalization to tensorial superspace
$\widetilde{\Sigma}^{(10|16)}$ can be associated with any of the
tensionless superstring actions (\ref{S(w)-0S}) with $w\not=0$.
The pure twistor form of the action similar to the
Berkovits-Siegel one for a supertwistor string is provided by Eq.
(\ref{S(w-tw)}). A possible drawback of this action is the lack of
a complex structure and hence of a $U(1)$ symmetry, which seems
relevant in applying the supertwistor string to Yang-Mills theory
\cite{Witten03,NB04,Siegel04}, although one cannot exclude the
(rather exotic) possibility of replacing this $U(1)$ symmetry of
the $w=0$ action by some other symmetry of the $w\not= 0$ models.
The same lack of complex structure results in a replacement of the
$SU(2,2|4)$ superconformal symmetry of the $w=0$ action by the
$OSp(8|8)$ generalized conformal symmetry of the $S(w\not=0)$
models (\ref{S(w-tw)})  ($OSp(4|8)$ for $w=1$).

The generalization of the twistor superstring to the more
conventional $D=10$ $N=1$ superspace is actually provided by the
tensionless superstring action (\ref{S(null)10}). To see this one
needs, following \cite{BZ91,BZ94}, to `extract the square root' of
the light-like vector $u_{\underline{a}}^{--}$ ({\it vector} Lorentz
harmonics) by introducing a set of 8 bosonic spinors
$v_{\underline{\alpha}}{}^-_q$ (basis of the spinor moving frame or
{\it spinor} Lorentz harmonics) highly constrained by
\begin{eqnarray}\label{sHarmonics}
 & 2v_{\underline{\alpha}}{}^{\!^{-}}_q v_{\underline{\beta}}{}^{\!^{-}}_q =
u_{\underline{a}}^{\!^{--}}\Sigma^{\underline{a}}_{\underline{\alpha}\underline{\beta}}
\; , \qquad v{}^{\!^{-}}_p\tilde{\Sigma}_{\underline{a}}v{}^{\!^{-}}_q = \delta_{pq}
u_{\underline{a}}^{\!^{--}}
 \; \\ \nonumber &
_{\underline{\alpha}, \underline{\beta} =1, \ldots , 16 \; , \quad
p,q=1,\ldots, 8\; }
\end{eqnarray}
($\Sigma^{\underline{a}}\widetilde{\Sigma}^{\underline{b}}+
\Sigma^{\underline{b}}\widetilde{\Sigma}^{\underline{a}}=
2\eta^{\underline{a}\underline{b}}$ and $\Sigma^{\underline{a}}$
was defined in (\ref{10DSigma})). Then, the action
(\ref{S(null)10}) reads
\begin{eqnarray}\label{S(null)10s}
S= & {1\over 8} \int\limits_{W^2} e^{++} \wedge  \Pi^{\underline{a}} \;
\tilde{\Sigma}_{\underline{a}}^{\underline{\alpha} \underline{\beta}}\;
v_{\underline{\alpha}}{}^{\!^{-}}_q v_{\underline{\beta}}{}^{\!^{-}}_q  \; , \qquad \\
\nonumber & \qquad \Pi^{\underline{a}} = dX^{\underline{a}}- i d\Theta
\Sigma^{\underline{a}}\theta\; , \qquad
\end{eqnarray}
and it is a clear counterpart of (\ref{S(BZ)}) but in terms of constrained spinors
$v_{\underline{\beta}}{}^-_q$, Eq. (\ref{sHarmonics}). It can be shown that these {\it
spinorial Lorentz harmonics} parameterize the celestial sphere $S^{8}$ represented as
the Lorentz group coset \cite{gdsghs}
\begin{eqnarray}\label{v-inS}
 \{v_{\alpha p}^{\;-}\} = {Spin(1,D-1) \over [Spin
(1,1)\otimes Spin(8)] \, {\subset}\!\!\!\!\!\!\times \mathbb{K}_{8}
} = \mathbb{S}^{8} \; ,
\end{eqnarray}
$\mathbb{K}_{8}$ being an abelian subalgebra. The $D=10$
counterpart of the original pure supertwistor form (\ref{TWS-S})
of the supertwistor string action can be obtained by presenting
the action (\ref{S(null)10s}) in the form
\begin{eqnarray}\label{S(null)10vm}
& S=  \int\limits_{W^2} e^{++} \wedge  (d\mu^{-\underline{\alpha}}_q
v_{\underline{\alpha}}{}^-_q  - \mu^{-\underline{\alpha}}_q
dv_{\underline{\alpha}}{}^-_q - i d\chi^-_q \chi^-_q) \, , \quad \nonumber \\ & {}
\end{eqnarray}
where the $D=10$ counterpart of the Penrose incidence relation reads
\begin{eqnarray}\label{twTvm}
\mu^{-\underline{\alpha}}_q= X^{\underline{a}}
\tilde{\Sigma}_{\underline{a}}^{\underline{\alpha}
\underline{\beta}} v_{\underline{\beta}}{}^-_q  - {i \over 2}
\Theta^{\underline{\alpha}} \,\Theta v^-_q \; , \qquad \chi^-_q =
\Theta^{\underline{\alpha}} v_{\underline{\alpha}}{}^-_q \; . \qquad
\end{eqnarray}
Due to the basic constraints (\ref{sHarmonics}), Eq. (\ref{twTvm})
results in $\mu^{-\underline{\alpha}}_q
v_{\underline{\alpha}}{}^-_p =
X^{\underline{a}}u_{\underline{a}}^{--}\delta_{pq} + {i\over 2}
\chi^-_q\chi_p^-$. This implies that Eq. (\ref{twTvm}), with
$v_{\underline{\alpha}}{}^-_p$ constrained by (\ref{sHarmonics}),
provides the general solution of the constraints
\begin{eqnarray}\label{SO(8)}
& \mu^{-\underline{\alpha}}_{[q} v_{\underline{\alpha}}{}^-_{p]} -
{i\over 2} \chi^-_q\chi_{p}^-=0 \; , \qquad \nonumber \\ &
\mu^{-\underline{\alpha}}_{(q} v_{\underline{\alpha}}{}^-_{p)} -
{1\over 8} \delta_{qp} \mu^{-\underline{\alpha}}_{p'}
v_{\underline{\alpha}}{}^-_{p'} =0  \; , \qquad
\end{eqnarray}
which play the role of the $D=4$ constraint (\ref{YY=0}). More
details on the twistor-harmonic formalism in $D=10$ and $11$ will be
presented in elsewhere.

Let us notice that the necessity of using constrained spinors to
describe the higher dimensional generalizations of the twistors
was recently noticed \cite{BP05} in the context of a two-time
physics generalization of the Penrose incidence relation, as well
as earlier in \cite{BZ91,BZ94}, in relation with the
generalization $ v_{\underline{\alpha}}{}^-_q
v_{\underline{\beta}}{}^-_q \propto p_{\underline{a}}
\Sigma^{\underline{a}}_{\underline{\alpha}\underline{\beta}}\;$,
$\,p_{\underline{a}} \delta_{pq} \propto
v{}^-_p\tilde{\Sigma}_{\underline{a}}v{}^-_q$ ({\it cf.}
(\ref{sHarmonics})) of the D=4 Cartan-Penrose representation of a
lightlike momentum, $p_{a}\sigma^a_{\alpha\dot{\alpha}}=
\lambda_\alpha \bar{\lambda}_{\dot{\alpha}}$, the other essential
ingredient of the Penrose twistor approach \footnote{Although in
$D$=6 and 10 one can use an unconstrained spinor
$\lambda_{\alpha}$ to build a light-like vector,
$p_a=\lambda\Sigma_a\lambda$ $\Rightarrow$ $p^2=0$, the matrix
$p_a\Sigma^a$ cannot in turn be expressed as
$\lambda_\alpha\lambda_\beta$.}. The twistor transform of the
tensionful $D=4$, $N=1,2$ superstring actions (\ref{S(BZst)}) has
been presented recently \cite{Uvarov06} (on the surface of
embedding equations
$\Pi^{\alpha\dot{\alpha}}v^-_{\alpha}\bar{v}^+_{\dot{\alpha}}=0$
and
$\Pi^{\alpha\dot{\alpha}}v^-_{\alpha}\bar{v}^+_{\dot{\alpha}}=0$).

To conclude, we mention that we did not consider in the present
context the interesting problem of the possible stringy origin of
the Yang-Mills current part $\int L_{G}$, Eq. (\ref{LYM-ff+}), in
the supertwistor string action (\ref{TWS-S}). In the light of the
discussion in Sec. \ref{Sect.Parent.tensionful}, it is tempting to
speculate that (\ref{LYM-ff+}) might originate from the heterotic
fermion contribution to the (tensionful) $D=10$ $N=1$ heterotic
string action. The main difficulty for such a scenario seems to be
the fact that the chirality of the heterotic fermions is opposite
to that of the fermionic coordinate function
$\Theta^{\underline{\alpha}}$, while the current generating
fermions in (\ref{LYM-ff+}) have the same worldsheet chirality
\cite{WS05} as the twistors in (\ref{TWS-S}) and the coordinate
functions $\theta^{\alpha}, \bar{\theta}^{\dot{\alpha}}$ in
(\ref{S-0S}).

\bigskip
\noindent {\it Acknowledgements}.

This work has been supported by research grants from the Spanish
Ministerio de Educaci\'{o}n y Ciencia (FIS2005-02761 and EU FEDER
funds), the Generalitat Valenciana (ACOMP06/187, GV-05/102), the
Ukrainian State Fund for Fundamental Research (N 383), and by the
EU network MRTN-CT-2004-005104 (`Forces Universe'). One of us
(C.M.E.) acknowledges the Spanish M.E.C. for his FPU grant. The
authors thank Dima Sorokin for useful comments on the manuscript.

\end{document}